# Combining Usability and Privacy Protection in Free-Access Public Cloud Storage Servers: Review of the Main Threats and Challenges

Alejandro Sanchez-Gomez[1,*], Jesus Diaz[1], David Arroyo[1]

*Departamento de Ingeniería Informática,*
*Escuela Politécnica Superior, Universidad Autónoma de Madrid*

## Abstract

The 21st century belongs to the world of computing, specially as a result of the so-called cloud computing. This technology enables ubiquitous information management and thus people can access all their data from any place and at any time. In this landscape, the emergence of cloud storage has had an important role in the last five years. Nowadays, several free-access public cloud storage services make it possible for users to have a free backup of their assets and to manage and share them, representing a low-cost opportunity for Small and Medium Companies (SME). However, the adoption of cloud storage involves data outsourcing, so a user does not have the guarantee about the way her data will be processed and protected. Therefore, it seems necessary to endow public cloud storage with a set of means to protect users' confidentiality and privacy, to assess data integrity and to guarantee a proper backup of information assets. Along this paper we discuss the main challenges to achieve such a goal, underlining the set of functionalities already implemented in the most popular public cloud storage services.

## 1. Introduction

The changes in data storage is one of the major characteristics of the evolution of information technologies [1]. It provides a clear example on the construction of abstract models to generate and manage information [2]. Certainly, in the beginning of modern communication networks, data value was very close to the physical nature of the data storage medium. In the 1950s, the access to Information and Communication Technologies (ICTs) was restricted to big companies and governmental organisations. As ICTs evolved, citizens were able to access and buy Personal Computers, which was concomitant with a popularization of the Internet. In this new scenario, most citizens could manage information assets through local data storage media as hard drives, floppy disks, CD-ROM, DVD, and USB-drives. Moreover, citizens and companies were able to make their information accessible to third parties through private and/or public web servers. Internet hosting services, in fact, are one of the most important elements configuring the digital transformation of business and society. Data outsourcing, nowadays, is a common practice, and more and more citizens and companies use Service Providers (SPs) to manage information assets. Consequently, either implicit or explicitly, users and companies trust third parties, i.e., the SPs, to handle information [3, 4].

On the other hand, the concretion of new interfaces to access data has an impact not only on the users of such interfaces. Indeed, the creation of new *modalities* to generate and control data has determined a paradigm change in software development. The increasing complexity of operating systems and the geographic distribution of resources have determined a *divide-and-conquer* methodology in software development. There is clear urge to decouple content management from data management, for the sake of agile development of new products and the efficient adaptation to changes in the production and/or exploitation environment [5].

Cloud data storage provides a whole set of solutions to ease user access to data. Moreover, it supplies an organisation with tangible competitive advantages: significant costs savings, improved degree of scalability, flexibility and resource pooling availability [6]. In addition, this new paradigm allows the creation of innovative software products by developers. Nowadays, cloud migration is one of the most important concerns for both organisations and individual users, since cost reduction is a must for every entity although it cannot imply losing efficiency and service quality [7]. Besides, before migrating into the cloud, it is necessary to realise and understand that it means to store their sensitive electronic assets into the providers' infrastructure, so this scenario introduces extra security and privacy challenges comparing to the traditional computing environment. Therefore, it seems necessary to design a process with which assure the security and privacy requirements of cloud users or organisations who want to migrate their assets into the cloud [6].

As stated above, in cloud environments data management is delegated to third parties according to different trust models [8, pp. 246-251], which represents concrete implementations of security requirements and expected functionalities. Actually, in cloud storage we have to distinguish between private, community, public and hybrid trust models [9]. Privacy threats differ depending on the type of cloud scenario and threats such as lack of user control, potential unauthorised secondary usage, or data proliferation are more dominant in public clouds [6].

Moreover, from the perspective of software developers, we should consider a more general landscape and not restrict ourselves to cloud storage and ponder over cloud computing. In cloud computing, a developer can opt for different trust models according to the so-called Software as a Service (SaaS), Platform as a Service (PaaS), and Infrastructure as a Service (IaaS), which constitute the SPI (Software, Platform, Infrastructure) model [10]. This trust ranges from a total control of cloud technologies (Infrastructure as a Service model), to a complete trust in the cloud SP (Software as a Service model). In this paper we analyse in detail this last trust model, underlying the main risks and solutions for end users' working with free public cloud storage solutions.

Public cloud storage is increasingly being adopted not only by domestic users, but also by enterprises. In specific, Small and Medium Enterprises (SMEs) use this kind of storage due to its high usability properties and as a means to reduce both production and management costs. This being the case, it would be very helpful for SMEs to have access to a guideline to properly adopt public storage without eroding security and legal expectations and/or requirements. This call is even more necessary when SMEs are dealing with free-access public cloud storage, since in many occasions this configuration is built upon a service provider that does not clarify properly how our data is treated [11].

This paper is focused on pinpointing the underlying security assumptions when we decide to use a certain free-access public cloud storage, but also on highlighting the means to enhance the protection of outsourced data. In order to face the security and privacy problems in free-access public cloud storage services, in Sec. 2 we summarise them by means of ten security challenges. We discuss the main concern of each of these challenges, and we provide a set of recommendations to tackle them. These recommendations are intended to identify cryptographic procedures and software solutions that can help both SMEs and end users to implement usable and low cost security solutions upon free-access cloud storage. Certainly, the proper combination of standard cryptographic measures and the functionalities provided by free cloud storages can lead to appealing serverless solutions. The set of recommendations explained in this work paves the way to enhance both the security and privacy protection of the end users of free cloud storage. However, not all these recommendations can be implemented by the cloud user. Some of them must be done by the Cloud Provider (CP). As a result, in these situations the cloud user should be aware whether her CP adheres or not to these good practices. In addition, we have to emphasise that the core contribution of this present work is on assuming that the developer has no control at all on the cloud storage platform. This is a key difference in regards to other surveys on cloud security [3, 12], since our effort is on configuring a guideline to develop client-side software for the secure and privacy-respectul adoption of cloud storage services. For this reason, in Sec. 3 we address the evaluation of the most popular free-access cloud storage providers. This study is very helpful to discuss and define the main requirements of those client-side software products. Furthermore, it configures a checklist to take into account in the risk analysis of the related cloud services, which would also be useful for SPs in the design of their platforms. Finally, we conclude this paper in Sec. 4.

## 2. Main threats and challenges in cloud storage

Over the last years, we have witnessed a lot of attacks and security holes in ICT[1]. Big companies have been compromised by hackers, which has derived in the erosion of their reputation and subsequent economical losses. These cyber-attacks have been possible due to the existence of some vulnerability or misconfiguration of the traditional procedures to secure information systems. Certainly, the deployment of information security systems is a very complex task. In today's scenario this is even more elusive.

Nowadays, mobility is a major fingerprint of ICTs. The increasing use of mobiles, tablets or VPNs, along with the huge quantity of data stored in the cloud, make really difficult to delimit a boundary with which a user could have a total control over her assets. As the boundaries of data storage are less defined, the same thing happens with the security perimeter [13]. When it is not clear the perimeter which we have to secure, we should consider how an organisation can protect itself against data leakage and malware. Therefore, it is clear that cloud customers never access the "real" network or hardware, since they work inside virtual constructs. As a result, the adequate evaluation of the security of cloud computing requires to define different models, different tools, and new foundations [14].

This new security perimeter, together with the huge rise in cloud adoption and the ever-decreasing cost of deploying applications and services in the cloud, have led to an increase of cyberattacks in the last years [15]. In fact, security breaches constitute the main concern that companies have in mind before migrating their business to this environment (see Fig. 1):

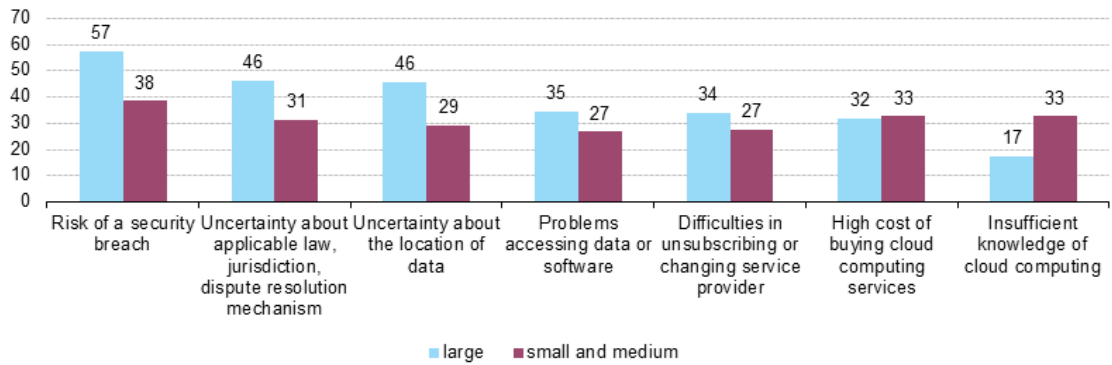

Figure 1: Factors limiting enterprises from using cloud computing services, by size class, EU28, 2014 (% enterprises using the cloud) (Image obtained from [16]).

---

[1]See http://www.informationisbeautiful.net/visualizations/worlds-biggest-data-breaches-hacks/ for a good summary of the most conspicuous cyber-attacks and their impact [Accessed 2016-06-04].

These users' concerns, and this high amount of attacks whose target is cloud storage services, have motivated us to present in the following subsections ten security challenges which could compromise the user's security when she uses public cloud storage. Thus, the main goal of this paper is to provide both companies and end users with a set of solutions in order to mitigate the well-known security issues in cloud environments.

### 2.1. Challenge 1: Authentication

The protection of information assets demands the convenient concretion of access control methodologies. These methodologies are mainly based on authentication and authorisation procedures. The authentication process consists of verifying the identity of a user requesting access to a certain information asset. In the context of cloud services, this process is usually performed by checking users' credentials so it must be done in a safe way. However, CPs often use basic authentication mechanisms, in which the user has to provide her credentials to her provider. Therefore, this user has to trust the fact that providers do not have access to her credentials and they also manage these credentials in a secure way. Hold Security, which is an Information Security consulting firm, shows that users should not rely on CPs: this company has recently recovered 1.2 billion stolen credentials that belong to different companies [17], showing a poor management of users' credentials [18].

During the last years, there have been several authentication problems in different cloud services. One of these is known as Man in the Cloud (MITC) attack [19], which has focused on file synchronisation services in the cloud, that allows individuals to store synchronised copies of files repositories in multiple devices. This attack is based on the following: when a user of a synchronisation application wants to make an initial connection to an account or to switch between accounts, it requires that the user provides her credentials (username and password) in an interactive way through the user interface. Since this moment, the synchronisation application relies on a synchronisation token, which is obtained after this authentication process, for further operations. This token is stored in a register or in a file in the user's computer, whose location is fixed and depends on each provider and it is also independent of the machine. Therefore, by copying this token from an account that is controlled by an attacker in the correct place in the victim's account, the attacker will be able to change the synchronisation application to the account that is represented by her token. Then, the attacker can gain access to the accounts of the victims without compromising their passwords. Without using the user's credentials in a explicit way, this attack cannot be detected by the owner of the account. We want to remark that this security hole has its root cause in an insecure implementation of the OAuth 2.0 protocol, which is designed for authentication and authorisation between applications [20, 21].

Therefore, the user has to make an effort and has to be sure to use a strong authentication mechanism in the cloud. This measure could help her to avoid security breaches like for example the incident called "Celebgate", which was on the last August of 2014, where almost 500 private pictures of various celebrities were posted on public web pages, which were obtained from iCloud [22].

*2.1.1. Solutions*

In order to face this authentication problem, we should used CPs that employ advanced password-based authentication techniques, like the Password-based Authenticated Key Exchange (PAKE) [23] or Secure Remote Password (SRP) [24] protocols, which effectively conveys a zero-knowledge password proof from the user to the server. With this last mechanism, an eavesdropper or man in the middle cannot obtain enough information to be able to brute force guess a password without further interactions with the parties for each guess. This means that strong security can be obtained since the server does not store password-equivalent data.

Furthermore, this verification process of user's identity could be better by adding Two-Step Verification (2SV) mechanisms, which is present in several free cloud storage services [25]. In Google Drive, the user could use Google Authenticator [26], which is a two factor verification mechanism that adds an extra layer of security to users' accounts. Dropbox has also an optional 2SV (via text message or Time-Based One Time Password apps) [27]. Box also provides the use of 2SV. In this case, the process requires a user to present two authenticating pieces of evidence when they log in: something they know (their Box password) and something they have (a code that is sent to their mobile device) [28]. Moreover, all OneDrive users can protect the login via One Time Code app or a text message [27].

On the other hand, in some situations in cloud storage, users are required to register using Personally Identifiable Information (PII) at a Cloud Provider (CP) to setup an account. Consequently, the provider can monitor the storage consumption behaviour of its users, i.e., every store, read, and delete operation can be tracked, and the provider thus has access to a complete behavioural profile of its users. Therefore, in this cloud services, anonymous authentication mechanisms are mandatory if the user really needs to get an adequate privacy in the cloud [29]. A way to achieve this type of authentication is through non-conventional digital signatures, like group signatures, which allow members of a group of signers to issue signatures on behalf of the group that can be verified without telling which specific member issued it [30].

Finally, we would want to point out some possible solutions to the MITC attack that we have mentioned in Sec. 2.1. One possible strategy could be, firstly, to identify the compromise of a file synchronisation account, using a Cloud Access Security Broker (CASB) solution that monitors access and usage of cloud services by the users; and secondly, to identify the abuse of the internal data resource, deploying controls such as Dynamic Authorisation Management (DAM) [31] around the data resources identifying abnormal and abusive access to the data [19]. However, we consider that the best way to solve this security issue is being very carefully in the OAuth implementation, using along with it a second authentication factor in order to have a high security level. To achieve this goal, CPs should be aware of the current potential weaknesses of this protocol, and to follow the well-known recommendations made by the cryptographic community about how to implement OAuth 2.0 in a secure way [20, 21, 32].

*2.1.2. Limitations*

In this and other scenarios, the deployment of security protocols imply a risk that has to be taken into account. As much effort used in the authentication process, we always can find a poor implementation of the cryptographic functionalities used, as it happened in the MITC attack, where an insecure implementation of OAuth 2.0 protocol comprised the security of several file synchronisation services. Heartbleed was also an important security hole [33], which was caused by an improper input validation (due to a missing bounds check) in the implementation of the TLS heartbeat extension, and it allowed to steal usernames and passwords from vulnerable sites. Another bug which had a significant relevance in the last January of 2015 was Ghost [34], that was caused by a buffer overflow in a system library, called glibc, that is used in Linux distributions. This bug allowed attackers to remotely take complete control of the victim system without having any prior knowledge of system credentials.

Regarding 2SV mechanisms, it has been shown they are not completely secure, since there have already been several situations in which an attacker could "break" this mechanism [35, 36]. In fact, in December of 2012, the Eurograbber banking Trojan had a significant impact. This trojan was used to steal more than 36 million Euro from some 30,000 retail and corporate accounts in Europe [37]. This attack infected the computers and mobile devices of online banking customers. Consequently, the 2SV mechanism was circumvented and the attackers were able to validate their illicit financial transfer [38]. Besides, if we are using a service from our mobile phone which includes 2SV sending a text message to this phone, the authentication process would be almost the same than if we only have to provide our username and password straightaway to this service [35].

On the other hand, we would like to add that nowadays it is widely used in several systems the so-called Single Sign-On (SSO) mechanism, which is a process that allows a user to enter a username and password in order to access multiple applications. Although it increases the usability of the whole system, these type of services, in which a token is used for authenticating a user, could be vulnerable in some situations to replay attacks [39]. As a result, the adequate combination of usability goals and security concerns is not easy to achieve, and the design of any procedure to improve usability should be thoroughly analysed before deploying the corresponding authentication process [40].

*2.2. Challenge 2: Information encryption*

Information encryption is an important aspect that users of free cloud storage should take into account. Certainly, the trust model and the risk model are very different when the SP is in charge of data encryption. If so, users are implicitly trusting SP, and this assumption can entail a threat to security and/or privacy [11].

With the current free cloud storage solutions, users have to completely trust their CPs, since although these providers usually apply security measures to the service they offer, such measures allow them to have full access to the data that they store [41]. Indeed, even though the data are encrypted in the CPs servers, the user should pose herself the question about who has the control over the cryptographic keys used for this encryption. It is not ideal that this owner is the CP, because in this case users would not have direct control on who can access their data.

Furthermore, providers could be also compromised because of hardware failures [42], the flaws of software [43], or the misbehaviour of system administrators [44]. Once one of these happens, a tremendous amount of data might get corrupted or lost [45].

### 2.2.1. Solutions

Solutions for protecting privacy require encrypting data before releasing them to the CPs, and this measure has gained a great relevance since Snowden's leaks [46]. Client-side encryption could be properly articulated to obtain privacy and data integrity protection.

Nonetheless, encryption can bring several difficulties in scenarios where querying data is necessary. In these particular situations, we could use fragmentation instead of encryption, so we only have to maintain confidential the associations among data values in contrast to the values themselves. This technique protects sensitive associations by splitting the concerned pieces of information and storing them in separate un-linkable fragments [41]. Moreover, from the point of view of the CPs, fragmentation is a possible optimisation, since they could detect redundant file blocks which would be impossible with data encryption [47].

### 2.2.2. Limitations

The information security requires the protection of its confidentiality, i.e., it is only accessible by only those who have permission. As stated at the end of the last subsection, data fragmentation is a possible solution for achieving this confidentiality. However, its main limitation is that it must be managed by the user rather than by the provider.

Cryptographic is another alternative to guarantee this confidentiality. When the custody of the information corresponds to third parties, the user has to decide who is in charge of encrypting the information. On the one hand, if this encryption is managed by the provider, the user's privacy could be compromised, as it was shown in Snowden's leaks. Certainly, if the user wants to have complete control over her information assets, client-side encryption seems to be a good option. Nevertheless, it implies that the user is in charge of keys generation and management [48], which could suppose a high workload for her, along with a security risk (e.g., the user could loss a cryptographic key and thus she could not access to her data). Thus, it is clear that it would be necessary to have a mechanism in order to face this problem.

In this context, some applications allow user to generate and manage her keys using for this external tools [49], to subsequently import them into the client application of cloud storage. Therefore, as long as this tool also follows the security recommendations of key generation, this option is the strongest, since it allows user to choose tools that were specially designed for key generation and in which she trusts [50]. The first security recommendation is to use random number generators, to the possible extent, to have solutions based on One Time Passwords (OTPs), or at least, with limited password reuse. The next step is to use key derivation functions like PBKDF2 [51], in which the OTP previously generated could be used to derive one or more secret keys. These two steps composed an ideal key generation process. However, some security breaches have been discovered recently in some key management tools, like in the case of LastPass [52]. For this reason, the user should not trust them and look for alternative solutions to avoid Single Point of Failure (SPOF) situations. Instead, the user could adopt the approach of distributed authentication, which consists of the generation of a strong secret from a password via interaction with two or more servers. With this new perspective, the user only has to know her password, and the compromise of one server exposes neither any private data nor any password [53]. Dyadic[2], a Multi-Party Computation (MPC) solution, meets with these requirements, since it protects secret keys and credentials by splitting them randomly between two or more servers. The unique property of Dyadic's technology is that all operations take place without ever bringing the key together, and so it is never in any one place to be stolen. Attackers need to simultaneously control multiple servers in order to learn anything, and this is made hard by using different operating systems, administrator credentials and more [54].

Apart from the key management problem, the user could lose some functionalities if she finally would decide to encrypt her data, since data encryption is not compatible with traditional deduplication mechanisms and she could not be able to make queries by keywords in the cloud [41]. However, the new paradigm called Secure Computation (SC) solves this problem, since it is essentially based on processing data that is protected by encryption [55]. In fact, the homomorphic encryption, which is a technology for programmable SC, is used by Dyadic for working with encrypted data. An alternative similar solution is Sharemind[3], which is also a MPC framework that can process encrypted data without decrypting it, using cryptographic methods which are costly (in terms of computing power) but efficient with the current technologies [56].

**Encryption of shared assets**

Finally, the sharing of documents over the cloud is another feature that could be even more difficult to achieve when we work with encrypted data. In this situation, it is necessary to share the encryption key of a encrypted file among the members of the group.

A first approach to solve this problem could be to make up a data packet based on the concept of the digital envelope container [57]. This packet will include the original document encrypted with a randomly generated symmetric key. Then, this key will be encrypted with each asymmetric public key which belongs to each member of the group and it will be attached to the packet. Therefore, when a member of the group who is not the owner of the document wants to open this document, first she would have to decrypt the symmetric encrypted key with her asymmetric private key, and then, she will be able to decrypt the document. However, the main limitation of this approach is the time that the owner of the document needs to build up the whole packet, and also the time that a member of the group have to spend to get the encryption key of the document. Besides, when a user decides to join/leave the group, all the documents that this group is sharing have to be updated to grant/remove the permissions to this user.

---

[2]<www.dyadicsec.com> Accessed 2016-06-04
[3]<http://sharemind-sdk.github.io/> Accessed 2016-06-04

Another approach that could solve this key management problem through a centralised cloud service [58], so that encryption keys are stored by the CP, and it acts as a central hub of communication with users that ask for keys. The main problem is that it makes the user to trust completely the CP, because it generates all the private keys, so it has the ability to decrypt all the data and communication in its domain. Furthermore, this solution is affected by a SPOF threat. Certainly, if the keys of the CP are stolen, then all communications are compromised; if it suffers from a denial-of-service attack, then it cannot distribute private keys [58].

The third approach includes key management through a trusted client-side authority [58]: users are segmented into populations called groups, each of which has read and write access to a different data partition. Data partitions within the cloud are encrypted so as not to reveal information to the CP. A trusted intermediary, called manager, is in charge of a group of users of any size, and it is responsible for all aspects of its security: it generates and holds all private keys, allowing read and write access to the data partition. The main advantage of this solution is that the manager is under the control of the client organisation, and ensures that key management functions need not be outsourced to an untrusted CP. Additionally, each manager handles the authentication of only a limited set of users that interact with its own data partition, so it reduces a possible overload.

The fourth and last approach is more advanced than the previous ones [59], and it proposes a framework that combines together into a protocol a proxy signature, with which the group leader can effectively grant the privilege of group management to one or more chosen group members; an enhanced Tree-Based Group Diffie-Hellman (TGDH), that enables the group to negotiate and update the group key pairs with the help of cloud servers; and a proxy re-encryption, that allows to delegate most computationally intensive operations to cloud servers without disclosing any private information. Therefore, this solution supports better the updating of encryption keys, because it transfers most of the computational complexity and communication overhead to cloud servers without leaking the privacy.

*2.3. Challenge 3: Inappropriate modifications of assets*

In cloud storage the data owner loses the control over her assets, which means a change about how the enterprise perimeter is built up. From a classical point of view, this perimeter was constructed upon the assumption of a fixed geographic location for the underlying ICT infrastructure, visible for the owner of all IT assets. This allows to protect the assets using several security controls, as firewalls, Intrusion Detection System (IDS), etc., which collectively forms the perimeter. However, in the cloud era, both location and data's owner are blurry and not clearly identified. Therefore, some of these security controls that are used in the old perimeter cannot protect IT assets in the cloud [13]. We can conclude that in this situation it is not enough to have a well-defined security perimeter. In fact, when the user uploads her assets to the cloud, she cannot be completely sure about when she later accesses her data again, they have not been modified by unauthorised third party, since an attacker can make small modifications in her assets that would be difficult to perceive. Even if the user has encrypted her data before uploading them to her CP, they can be modified and the user probably will not notice it.

On the other hand, for security considerations, previous public auditing schemes for shared cloud data hid the identities of group members which shares these data. However, the unconstrained identity anonymity will lead to a new problem, that is, a group member can maliciously modify shared data without being identified. Since uncontrolled malicious modifications may wreck the usability of the shared data, identity traceability should also be retained in data sharing [45].

### 2.3.1. Solutions

The first solution that the user could have in mind is to perform a hash function in each file that she wants to upload to her CP, so then, when she accesses to these files, she could perform again this hash function over the downloaded file, and verify if it is the same as the first hash calculated. However, if this user has to deal with a big quantity of files, she could find the problem of where she can store all these hashes.

Another approach for solving this problem that the cryptographic community proposed was an online mechanism to verify both data availability and integrity, that consists of inspecting a small portion of these data combining a cryptographic processing. These methods are called retrievability proofs, which seems to have been first defined by Juels and Kaliski [60]. Their idea was to encrypt the information and then insert data blocks called "sentinels" on the encrypted information in random points. Sentinels are generated using a cryptographic function whose values cannot be distinguished from the encrypted data without knowing the function's parameters. In this way, a file is prepared to be sent to the cloud server to store it. The provider only sees a file with random bits, so she is not able to distinguish between what parts of data are original and what parts have been inserted randomly (sentinels). The data owner can verify if her data are intact, asking for a random selection of the sentinels that are returned by specifying the positions that only the owner knows. As the provider does not know where the sentinels are, any change in data can be detected. However, this solution supposes an overload for the user who has only limited computing and network resources, because she would have to be in charge of verifying from time to time the integrity of her data stored in the cloud.

At this point, we should distinguish between content integrity (achieved with hash, CRC, sentinels, etc.) and both content and source integrity, which can be achieved with specific cryptographic methods. Therefore, we can agree that an exhaustive verification protocol is necessary in order to guarantee both data coherence and data integrity. As we have already seen, integrity verification is really important in all cloud computing schemes where the user does not trust provider's service [61]. Although it is possible to reach a conclusion about physical identity, it is not easy to establish a relationship between physical identity and digital identity. Cryptography provides the means for associating a digital identity to a user through asymmetric cryptography [30]. Consequently, the verification of the integrity of IT assets has been conceived and managed through conventional digital signatures.

However, in modern IT networks conventional digital signatures do not offer the whole of set of features that are required. In some situations, for example, it is necessary to design a signature scheme in which the message or document has to be signed by multiple users (e.g., if we are dealing with a committee that must sign the whole document). On this ground, we have to take into account that there exist over 60 digital signature models [30], whose classification is not easy, as they have a wide set of different characteristics. This being the case, group signatures could be used as anonymous authentication methods in the cloud. On the other hand, multi and aggregate signatures could be used when a file is digital signed and shared between several users in a cloud storage service, avoiding to produce multisignatures that are of linear size in the number of participants and with linear verification time (also depending on the participants). Finally, identity-based signatures could eliminate the need of distributing public keys in the cloud, allowing the verification of digital signatures just from the identity that the signer claims to own [30]. The transition from the need of communication networks and the features provided by these different schemes is not an easy task. In fact, this connection should be guided by strict evaluations of the theoretical bases.

Moreover, as we have seen at the end of Sec. 2.3, the auditing of shared cloud data demands the traceability of users' operations. However, this traceability could imply a risk in terms of users' privacy. Therefore, it is required an efficient public auditing solution that can preserve the users' privacy and, simultaneously, guarantee the identity traceability for group members. In [45] it is proposed an scheme in which a group manager is introduced to help members to generate authenticators and protect users' privacy. More specifically, identity traceability is achieved by two lists that record members' activity on each data block. Besides, the scheme also provides data privacy during the authenticator generation by utilising a blind signature technique.

### 2.3.2. Limitations

Digital signatures are often performed using the asymmetric cryptographic algorithm RSA. The main disadvantage of this algorithm is the time required for its key generation, although it usually has to be done just once. Nevertheless, we could improve this time using elliptic curves [62], which provide the same security as RSA but with smaller key size. In fact, RSA key generation is significantly slower than ECC key generation for RSA key of sizes 1024 bits and greater [63, 64]. In addition, the cost of key generation can be considered as a crucial factor in the choice of public key systems to use digital signatures, especially for smaller devices with less computational resources [65]. Nonetheless, in some cases it is not necessary to generate RSA keys for each use. For this situation, we would remark that the problem mentioned before is not so dramatic, since RSA is comparable to ECC for digital signature creation in terms of time, and it is faster than ECC for digital signature verification [66].

Furthermore, we could incorporate Merkle trees [67], such that we will reduce the number of digital signatures to handle. Actually, as a real example which uses Merkle trees we have the well-known system Bitcoin[4].

## 2.4. Challenge 4: Availability

Users of cloud environments typically want to place a large amount of data in cloud servers. Some of this information might not be used during long periods of time, but it must be available when required. CPs check data availability by storing files with redundancy or error correction. Although the main cloud storage providers deploy hardware redundancy mechanisms, from a general point of view, users cannot be sure whether their files are vulnerable or not against hard drive crashes [68].

### 2.4.1. Solutions

Despite high adoption rate among consumers, cloud storage services still suffer from many functional limitations and security issues. Recent studies propose the utilisation of RAID-like techniques in addition to multiple cloud storage services as an effective solution [69]. Therefore, as a simple solution, the user could think about splitting her data between different CPs, that is, saving a copy of each of her files in each CP. In this context, if one of these CPs would have a critical problem in its servers, the user could access to all her files stored in other providers. However, the management of files stored in multiple cloud storage servers is not an easy task.

Fortunately, recently some cloud storage managers has been proposed to handle cloud storage in multiple SPs [70]. Moreover, in [69] it is proposed a solution for mobile devices that unifies storage from multiple CPs into a centralised storage pool, which represents a clear improvement in terms of availability, capacity, performance, reliability and security.

### 2.4.2. Limitations

The solutions presented in [70] are private, so the user would have to pay a fee each month if she would want to use this service without limitations. Therefore, it is for this reason that these solutions have no place in our "free" cloud storage solutions mindset. Additionally, using these proposals the user could not feel comfortable, since she would let another third-party company manage her files. In this case, the user would have to read each company's term of services to see how they manage her files and she would have to trust them. This multi-cloud solution could be implemented on the cloud user side.

---

[4]https://bitcoin.org/en/ Accessed 2016-06-04

*2.5. Challenge 5: Data location*

The location of user data in the cloud can be a critical security issue. When the provider guarantees that the data are stored within specific geographic area, users might not have the assurance about this fact. On this point, it had a great relevance the decision of the European Court of Justice made on the last October of 2015: the annulment of the EU-US data sharing agreement named Safe Harbour [71]. This revocation prevented the automatic transfer of data of European citizens to the United States of America (USA). However, since the last February of 2016 there is a new framework that protects the fundamental rights of anyone in the EU whose personal data is transferred to the United States [72]. These agreements prove the huge legal complexity of the ubiquity in cloud storage systems.

*2.5.1. Solutions*

Cloud service latency is a hint to infer the geographic location of data as stored by CP. Indeed, it will take more time in their transmission as they are farther from user location. These measures have to be carried out with high degree of precision since the information moves really quickly in electronic communications.

An example of location proof is the use of distance bounding protocols [73]. These protocols always imply a timing phase in which the verificator sends a "challenge" and the provider responds. The provider is required to respond within a time limit which depends on the distance between the provider and the user. Cryptographic authentication procedures are used to ensure that the responses come from the intended provider instead of other entity that says to be the provider.

In the cloud, this method can be used to verify with accuracy the continent, and even the country where data are stored.

*2.5.2. Limitations*

In the proposed method there are several uncertainties that can complicate this measure, like a delay in the Internet communication. Furthermore, there can be some delays caused by the time that the server spends in order to access the disk and to the correct sector where the blocks of data are stored. In fact, this measure is really dependent on the storage service, since for example Dropbox has to decrypt the information before sending it (as the encryption/decryption is made in the server), whereas this is not necessary in Mega, since this operation is made in the client. Finally, we want to remark that this method would be deployed on the cloud user side.

*2.6. Challenge 6: Data deduplication*

With the huge potential storage space offered by CPs, users tend to use as much space as they can, so providers constantly look for techniques which can minimise redundant data and maximise free space available. One adopted technique is deduplication. The simple idea behind deduplication is to store only once deduplicated data (both files and blocks). Therefore, if a user wants to upload a file (block) that is already stored, the CP will add the user in the list of owners of this file (block). Deduplication is able to save space and costs, so that many CPs are adopting it.

However, the adoption of deduplication is not an straightforward decision for a CP. The CP must decide between file or block level deduplication, consider server-side vs. client side deduplication, and opt for either single user or cross user deduplication [74]. Some of these possibilities imply the existence of side channels that pose privacy matters [75], which makes some users to demand data protection and guarantees of confidentiality through encryption. Unfortunately, deduplication and encryption are two technologies that come into conflict between them. While the objective of deduplication is to detect data segments that are identical and to store them only once, the result of encryption is to make indistinguishable two identical data segments. This means that if data are encrypted by the user in a standard way, the cloud storage provider could not apply deduplication since two identical data segments will be different after their encryption. On the other hand, if data are not encrypted by users, the privacy cannot be guaranteed and data will not be protected against *curious* CPs [76].

### 2.6.1. Solutions

Several solutions have been proposed to mitigate different deduplication privacy concerns [75, 77]. One of these solutions, that can be used for solving these two conflicting requirements mentioned above, is the use of Convergent Encryption (CE) [78]. This technique consists of using as encryption key the hash of the content of data we want to encrypt, such that two equal files will generate the same encrypted file. This will allow both deduplication and assets encryption [79].

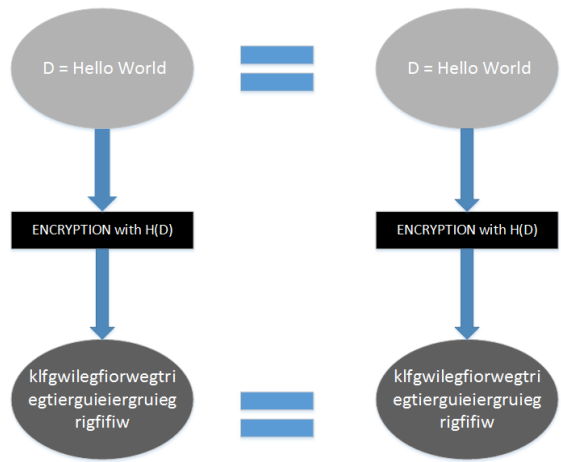

Figure 2: Example of convergent encryption, producing the same encrypted block encrypted from different clients.

Apart from this solution, there are other alternatives which may solve the problems aforementioned, like the ClouDedup tool [76]. This solution proposes the inclusion of an intermediate server between the users and the SPs in the cloud. Users will send to the intermediate server their blocks encrypted with CE, along with their keys, so that the server is able to identify duplicate blocks between all blocks which are sent by different users. Afterwards, the server will be in charge of deduplication and management of a second non-convergent encryption which confirms the security of the process, sending to the storage service only blocks that are not duplicated.

Finally, as we stated above, CPs can save storage and space with file deduplication. However, users do not directly benefit from these savings. On this point it would be useful to consider proposals as ClearBox [80]. Clearbox is a novel storage solution which allows a CP to transparently prove to its customers the deduplication percentages of the encrypted data that it is storing. Therefore, ClearBox provides cloud users with a means to verify the real storage space that their data is occupying in the cloud; this allows them to check whether they qualify for benefits such as price reductions, etc.

### 2.6.2. Limitations

First of all, CE suffers of several weaknesses, including dictionary attacks. Furthermore, it adds a privacy risk: with CE in the client side, a malicious user could know if it already exists some particular information in the cloud, which could be not acceptable for some users. However, one possible solution that could counteract the weaknesses of CE is taking into account the popularity of data segments. The data segments which are stored by several users, that is, the most popular, are protected by the weak mechanism of CE while unpopular data segments, which are unique in the storage, are protected through symmetric encryption scheme using a random key, which provides the highest level of protection while improves the computational cost in the client. This approach allows an efficiently deduplication since popular data segments that are encrypted with CE are the one that need to be deduplicated. This scheme also assures the security of data, since sensitive data, less popular, have a great protection with semantically-secure encryption while popular data segments do not suffer the weaknesses of CE since they are less sensible as they are shared by several users [81].

Nevertheless, this approach has another challenge: users have to decide about the popularity of each data segment before its storage and the mechanism with which this decision is taken leads the way for a serial of revelations that are really similar to the CE weaknesses. The focus of the schemes based on popularity is on the design of secure mechanisms in order to detect the popularity of data segments [81]. Thus, in this schema, data need different levels of protection depending of their popularity, so a data segment could be popular when it belongs to more than t users (where t is the popularity threshold). The popularity of a block is seen as a trigger for its deduplication. In the same way, a data segment would not be considered popular if it belongs to less than t users. This is the case of highly sensitive data, which are unique and they do not have to be deduplicated. When data segments that are unpopular begin to be popular, this is, threshold t is reached, the encrypted data segment is converted in its form of convergent encrypted in order to enable its deduplication. Therefore, when a user wants to upload a data segment, she must first discover the popularity degree of data in order to perform the appropriate encryption operation.

Finally, as far as we know, the best solution proposed is the tool called ClouDedup [76], which in its architecture, apart from the basic storage server, has a metadata manager and an additional server. This last server adds an additional non-convergent encryption layer on the most sensitive data in order to prevent well-known attacks against CE and to protect data confidentiality. On the other hand, the metadata manager is responsible of the key management task, since block-level deduplication requires to memorise a large number of keys. Therefore, deduplication is performed at block-level and an efficient key management mechanism is defined in order to avoid that users have to store a key for each block.

### 2.7. Challenge 7: Version control of encrypted data

If the upload of encrypted files to the cloud is allowed, there exists the problem of the version control of encrypted data. Version control of plaintext files is trivial, since it is possible to check which part of the original file has been modified. In information encryption, it is a must to use a secure cipher mode as for example CBC [82], in which an isolated modification in the original file modifies the whole encrypted text produced (condition that is essential in all cipher methods). However, these cipher modes make that the version control system cannot be able to know what specific part has been modified, having to store each file several times, one for each file version. This increases the space that is needed for the repository and reduces the efficiency.

#### 2.7.1. Solutions

An approach that has been proposed recently is to split the files uploaded to the cloud in data objects [83]. Each data object has a variable size with a maximum configurable size. If one file has a smaller size than this maximum, then it can be represented by only one object; otherwise, it has to be divided into multiple objects. As a result, if some modifications are produced in a big file, the user would only upload the objects that have been modified, instead of the whole file, saving costs on both assets upload and storage.

In many cases, the same object can appear in multiple backup versions, or different objects can have the same content in the same or different versions. Therefore, if two objects have the same content, then it is only necessary to store one object in the cloud and to create small pointers in order to reference these objects. For determining if two objects are identical, hash functions can be applied.

In this fashion, the user could place different backups at different times, and organise these backups in different versions. For each version, there would be a metadata object that describes data objects. Fig. 3 shows how different backup versions are uploaded. Suppose that at time $t_1$, we want to upload the version $V_1$ that has four objects: ($O_1$, $O_2$, $O_3$ and $O_4$). Imagine that after, in the instant $t_2 > t_1$, $O_1$ and $O_2$ are not included, and $O_5$ and $O_6$ are added. Therefore, the new version $V_2$ is going to upload physical copies of $O_5$ and $O_6$, and its metadata objects will have pointers that reference to the physical copies of $O_3$ and $O_4$ in the version $V_1$. Finally, all metadata objects and data objects are stored in the cloud. This being the case, it would have a different encryption key for each data block. Consequently, if a modification on the version $V_1$ is done, it is only required to upload the block that has been modified.

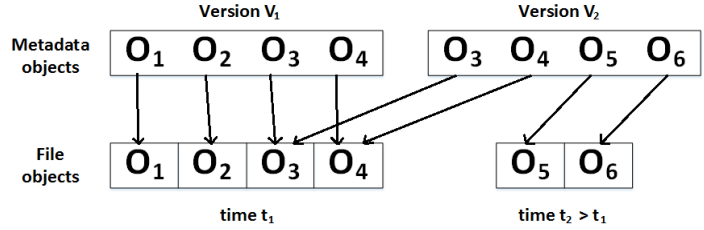

Figure 3: Control version mechanism

We would also like to add two tools which allows version control over encrypted data. The first one is SparkleShare[5], a solution based on a git repository that provides easy server-side encryption. Git-crypt[6] is another project which enables transparent encryption and decryption of files i a git repository. In [84] Git-crypt and Git-encrypt are thoroughly analysed in comparison with the default Git implementation. The authors conclude positively in regards to the expected functionality from a version control system, and also with respect to the privacy protection. Therefore, both tools enable the use of a non fully trusted Git server.

*2.7.2. Limitations*

The solution which consists of dividing the file in different chunks could be a good option. However, it has the problem of higher management costs for the user, since she has to be in charge of splitting the files she uploads and to compare each block in order to perform an appropriate version control. Moreover, in case of applying this solution, the main problem would be the management of a large number of encryption keys, one for each data object stored in the cloud [83].

In addition, there are some weaknesses [85] related to the two tools mentioned at the end of the last subsection. On the one hand, SparkleShare does not provide any key change mechanism; the encryption password is saved on each SparkleShare client as plain-text and can easily be read by everyone; and the filename is not encrypted at all, so an attacker can easily see what files you have stored. On the other hand, in Git-crypt, the filename is not encrypted, and the password can be found in plaintext on the client's computer. As a matter of fact, the management of metadata in version control system can imply some security risk, even if one uses solutions as Git-crypt or Git-encrypt [86]. In addition, key exchange is not an easy process and it is still an issue to be solved in coherence with usability criteria [84].

---

[5]www.sparkleshare.org Accessed 2016-08-15
[6]https://github.com/AGWA/git-crypt Accessed 2016-08-15

## 2.8. Challenge 8: Assured deletion of data

The assured deletion of data guarantees that files are inaccessible once user has requested to CPs to delete them. It is not desirable to keep data backups permanently, since sensitive information can be exposed in the future due to a security breach or a poor management made by CPs [83]. Moreover, CPs can use multiple copies of data over cloud infrastructure to have a great tolerance against failures. As providers do not publish their replication policies, users do not know how many copies of their data are in the cloud, or where those copies are stored. Hence, it is not clear how providers can delete all the copies when users ask for removing their data. In the particular case of Dropbox, when a user removes a document on this platform, this file is queued for permanent deletion. Queuing is used to enable file recovery and version history [87], but it could enclose a security risk.

### 2.8.1. Solutions

There are different ways to get assured deletion of data, but the best approach is to use client-side cryptographic protection, since if the user destroys the keys which are necessary to decrypt her data blocks, then these blocks will be inaccessible [83].

### 2.8.2. Limitations

The cryptographic protection can have a main limitation, since we must assume that decryption keys are maintained by a key management system that is totally independent of the cloud system and it is totally controlled by users, which is a great overload for the latter. Therefore, the same limitations explained in Sec. 2.2.2 can also apply for this section. On the other hand, note that client-side encryption only protects users' data against CPs, but not against changes in the users groups in which is performed a key distribution protocol. In this specific case, if a user is revoked from a group, she could still access all previous versions of shared files that belongs to this group if she has previously stored them locally. However, the best advantage of this solution is that it is implemented on the client side.

## 2.9. Challenge 9: API's validation

The current technological scenario has been called by many as the era of containers (*container age*), since a large number of initiatives are based on cloud solutions built by third-party software products. Thus, it has emerged a trust model that is not always taken into account when analysing the security risks of our proposals. Indeed, many of these third-party solutions are given by APIs that have not always been properly validated in terms of security [88],[89].

*2.9.1. Solutions*

Due to the high complexity of the current information systems, first, is highly recommended the use of a Secure Development Lifecycle (SDL). This methodology provides best practices for developing secure applications, with the aim of overcoming the challenges of making software secure [90]. SDLs are the key step in the evolution of software security and have helped to bring attention to the need to build security into the Software Development Life Cycle (SDLC).

Along with SDLs, since there are several details to take into account in a software development, developers should rely on automatic tools. Indeed, the application of formal methods to verify security properties has received more repercussion lately in order to avoid this type of security holes [91]. Several methodologies support the application of these mechanisms and theories with the aim of get high security guarantees for critical systems and protocols. However, a straight application of these tools can involve a high consumption of resources during the early stages of design, since some faults can be detected by simple performing an informal analysis [92].

In this aspect, an intelligent approach that a developer could take is to add an extension over some public cloud storage services which already exist, in order to counteract the weaknesses that these services have. Then, she will only have to validate her extension, since these public cloud storage services have already been validated by the security community.

In this guise, a methodology for the design of secure protocols is proposed in [92], which covers the whole design process and takes into account the problems mentioned before, following an iterative approach that consumes fewer resources in the early stages than in the following stages. As a consequence, the complexity of the failures detected increases as we progress through the methodology steps. Furthermore, a correct abstraction of the protocol's environment and attacker model are made. In this vein, the methodology proposed allows a fine control in the attacker modelling in order to fit it to the specific context in which the protocol is going to be executed.

Therefore, we have to remark the necessity of an adaptive process with feedback when any protocol of information security is designed. Any possible methodology must be built over evaluation tools for the objectives and acceptances which are involved in design and implementation stages, and any threat identified during the implementation or production should be managed as feedback for future improvement in the design of the system [93].

On the other hand, along with tools as Maude-NPA [94] and the framework STRIDE [8], we have to highlight companies as Cryptosense[7], which is a recent start-up which is looking to commercialise techniques for analysing the security of APIs [93].

---

[7]<https://cryptosense.com/> Accessed 2016-06-04

*2.9.2. Limitations*

Nevertheless, the human factor also takes part in the analysis of protocols procedure, so a bad formalisation can make that some errors are not detected. Then, as in any engineering process, the solution presented does not provide the 100% of success, but it helps avoid a great quantity of failures.

Finally, assuming that the CPs are validated by the cryptographic community, the solutions proposed for this challenge will be used to validate the new developments made by cloud developers. Nevertheless, the cloud users should verify that their cloud providers follow these good practices.

*2.10. Challenge 10: Usable security solutions*

Daily activity of today citizens is very dependent on cloud services. Nevertheless, this use is not always a secure as it should be. The incorporation of security measures is consequently required, but the related changes should not erode users' acceptance of the so-modified cloud services [95, 96]. In other words, secure cloud services solutions should be also easy to use.

*2.10.1. Solutions*

The so-called security-by-design and privacy-by-design paradigms [97] are hot topics in cryptographic engineering. On this point, we have to consider standards [98], security reference architectures [99], and well-known technologies as a basic framework for designing new security systems.

Furthermore, after the security solution is developed and before it is published for its use, its usability has to be tested. For example, in [95] it is proposed to evaluate a specific security solution, PGP 5.0, through a cognitive walkthrough, a heuristic evaluation and a laboratory user test, whose combination produces an exhaustive evaluation. In [100] Human-Computer Interaction (HCI) methods and practices are provided for each phase of the SDLC, in order to create usable secure and privacy-respectful systems. Moreover, there are recent efforts on formalising and automatically assessing the role of human factors in security protocols [101, 102].

*2.10.2. Limitations*

Firstly, a key limitation to consider is that security and privacy are rarely the user's main goal, and they would like privacy and security systems and controls to be as transparent as possible [100]. On the other hand, users want to be in control of the situation and understand what is happening. As a consequence, these two factors make harder to develop usable security applications.

Finally, in order to be successful in the development of usable security solutions, it is needed to have a multidisciplinary team including experts from the fields of IT, information science, usability engineering, cognitive sciences and human factors. This could suppose a main limitation, since no all companies have enough resources in order to form this kind of heterogeneous work group.

In conclusion, the solutions proposed for this challenge are for the developer who wants to implement new security developments over the current cloud platforms. However, it is recommendable for the cloud user to verify that her CP also uses the methodologies presented.

## 3. Security and privacy in the most popular free-access cloud storage servers: a very first analysis

Once we have underlined the main risks we can find when locating our data onto free-access cloud servers, the next step is to discuss which is the actual situation in some of the most popular solutions for this kind of cloud service. However, this is not an easy task, since the involved service providers do not offer the required information to conclude about the fulfilment of security and privacy goals (see Appendix B for a summary of the main characteristics of the CP under consideration). As a result, we have conducted a first approximation to the problem in order to underline the convenience of creating a guideline on the basis of the above security challenges.

According to different reviews published in [103, 104], Dropbox, Google Drive, OneDrive, Box, and Mega are the most popular free cloud storage solutions in the current market. In this section we are going to verify the security challenges presented above in each one of these five cloud storage services. This analysis is going to be focused on those challenges concerning the CP. Certainly, it is possible to distinguish those challenges that are connected to requirements to be satisfied either by the CP or the cloud user. Moreover, some of the solutions provided in the different challenges resort to solutions including additional servers, i.e., servers that are not managed by the CP but by the user. Taking into consideration this matter, we have classified the solutions for each challenge in Table A.1. As this table informs, CP are the main entity on charge of authentication, information encryption, availability, data deduplication, version control, API's validation, and the implementation of usable solutions. Since it is not possible to obtain all the information regarding these points, we are going to discuss the most relevant ones. The following is based on the main information extracted from the CPs' sites and online reports on cloud storage. For the sake of completeness, this information is included in this work as an appendix (Appendix B).

## 3.1. Challenge 1: Authentication

Authentication in the most popular free cloud storage services is mainly based on password-based mechanisms. Although registration and authentication is carried through a secure TLS channel, the hashed passwords are stored in the CP and this could endow a security risk. In fact, there are some evidences about the leakage of hashed passwords in services as Dropbox [8]. Therefore it is advisable to use multifactor authentication. As we can read in [105], Google Drive is the CP with a larger set of possibilities to backup two factors authentication. On the opposite side we have Mega, which does not support multifactor authentication.

Although Mega suffers from the lack of multifactor authentication procedures, it is necessary to pinpoint that Mega is the only one CP that does not store the users' password neither hashed or in clear. In the Mega registration process, the user's email is encrypted with an AES key derived from the password introduced by the user as an authenticating token. An AES master key is created, which is encrypted with the user's password, along with a RSA key pair that is encrypted with the master key. Both keys are then sent encrypted to the server. When the user logs in, this token is regenerated from the email address and the password introduced by the user. As a result, Mega is secure against any possible leakage concerning users' credentials.

## 3.2. Challenge 2: Information encryption

As we have stated in Sec. 2.2.1, the proper protection of users' privacy in the cloud calls for the application of client-side encryption. In the case of the most popular free cloud storages, Mega is the only solution enabling client-side encryption. In Mega, the files are encrypted on the client's side, using 128-bit AES that are generated randomly. These keys are also encrypted with the user's AES master key, which is encrypted again with an AES key derived from the user's password [50].

In the other platforms the information in transit is protected by SSL, and most of the solutions are also protecting the information at rest. However, in the case of OneDrive encryption at rest is only available for business users[9]. All the CPs here considered use AES-256 to encrypt the information stored in their servers[10,11], with the exception of Google Drive that applies AES-128 to protect the data[12].

Key management is a critical matter in the context of server-side encryption, due to both security and legal/normative compliance. Indeed, security cannot be guaranteed unless the secret keys of encryption are not highly protected. On this point, for example, Dropbox sustains that only a small number of employees have access to the keys used to encrypt the information[13]. Moreover, the management of secret keys should be further protected by enabling Hardware Security Modules (HSMs). HSMs are tamper resistant hardware solutions in charge of secret keys. Although HSMs are an adequate solution for key management, we have to be aware of any possible information leakage in the extraction of secret keys from HSMs during the encryption/decryption process. In fact, there have been reported security problems in some solutions as the Box's Enterprise Key Management procedure[14].

---

[8] https://haveibeenpwned.com/ Accessed 2016-10-09
[9] https://technet.microsoft.com/en-us/library/dn905447.aspx Accessed 2016-08-23
[10] https://www.dropbox.com/security Accessed 2016-08-23
[11] https://community.box.com/t5/Account-Information/Is-My-Data-Encrypted/ta-p/32 Accessed 2016-08-23
[12] https://apps.google.es/intx/en/faq/security/ Accessed 2016-08-23

Regarding normative compliance, if we are pondering to outsource data with personally identifiable information we should know how the CP is handling the secret keys. To clarify this point, let us take into account health data and the HIPAA normative [106]. This normative demands a series of requirements and the sign of a Business Associate Agreement (BAA). Although Google Drive, Box, and One Drive have signed a BAA, this is not the case of neither Dropbox nor Mega[15].

### 3.3. Challenge 4: Availability

The reliability of CPs is another key point to evaluate the convenience to migrate our information assets to the cloud. Most of the free cloud storages assert that their service is always up. Nonetheless, some studies have shown that the most popular cloud storage services suffer downtimes[16]. Although there are not recent works analysing the consequences of the outages of free cloud storage services, the previous evidences advice for the deployment of multicloud solutions.

### 3.4. Challenge 6: Data deduplication

In terms of security and privacy, it is highly relevant to know the main aspects of both single-user and cross-user deduplication. This information is not easily accessible. Nevertheless, we can examine the case of Dropbox and Mega.

In Dropbox, there were some evidences about an insecure implementation of data deduplication[17]. Dropbox reacted and solved this security problem [107]. However, it is not easy to confirm this fact and, consequently, we should consider deduplication in Dropbox as a potential risk for the information assets that we store in this platform[18].

On the other hand, Mega carries out both client side single-user deduplication at a file level and server side cross-user deduplication of the files once they are encrypted [50].

---

[13] https://www.dropbox.com/help/27 Accessed 2016-08-23
[14] https://www.sookasa.com/blog/boxs-enterprise-key-management-solution-falls-short-on-security/ Accessed 2016-10-09
[15] https://adeliarisk.com/hipaa-compliant-cloud-storage/ Accessed 2016-10-09
[16] http://royal.pingdom.com/2012/06/21/cloud-storage-shoot-out-google-drive-vs-dropbox-vs-skydrive-vs-box-com/ Accessed 2016-10-09
[17] https://github.com/driverdan/dropship Accessed 2016-10-15
[18] http://webapps.stackexchange.com/questions/54633/does-dropbox-no-longer-de-duplicate-files-across-users Accessed 2016-10-15

*3.5. Challenge 8: Assured deletion of data*

Deleted data in the most popular free cloud storage is placed in the recycle or trash bin. The erased information is not deleted completely once the deletion operation has been executed by the cloud user. Certainly, this data is kept by the CP for 30 days in all cases. In the case of Mega, the deleted assets are moved to the recycle bin. However, Mega does not specify in the terms of its service what it does with a file once it is deleted from the bin.

Taking into account the previous comments, if we want to be sure that others cannot access the deleted information, then we have to use client-side encryption upon the service provided by the free cloud storage platform. This being the case, Mega is the safest option on this matter.

## 4. Conclusions

Nowadays, cloud storage is a relevant topic due to the increase on the number of users who place their assets onto the cloud. However, these users often do not trust about where their data are going to be stored and who is going to have access to this data. It is for this reason that many users feel the obligation of applying security measures in order to have a total control over their data. More concretely, the user could find authentication, integrity, availability, confidentiality, and privacy problems. In the specific case of enterprises these imply key considerations that should be included in any cloud service agreement.

These concerns have motivated us to identify in this paper the most relevant security problems that users of free cloud storage services can find. For each identified security challenge, we have outlined some solutions and limitations, keeping always in mind that some of them can only be implemented on the CP side. Since cost reduction is a major concern in the case of SMEs, in this work we have highlighted those security solutions that can be implemented using serverless architectures. Correspondingly, we have discussed standards of information security that can be combined with the functionalities of free cloud storage services according to a serverless architecture. Table A.2 summarises all the solutions proposed in this paper, along with their limitations and suitability for a serverless setup.

After discussing the main security challenges in cloud storage, we have conducted an analysis of the security and privacy concerns in the most relevant free public cloud storage solutions. This information is very important for the users of cloud storage, specially for the enterprises that must satisfy legal or normative requirements. Furthermore, it is useful as guideline for software developers interested in the design of secure low-cost cloud storage solutions according to specific requirements. Actually, the methodology proposed in this paper helps in both the identification of these requirements and the evaluation of the risks associated to their adoption.

Finally, we have to take into account that the world of security information evolves really quickly every day, so that we have to be aware with the new potential security problems which could affect the security and privacy of the user in cloud storage environments. In this evolving scenario, our work is intented to help cloud users to evaluate the cloud service agreements according to the recomendations of the new ISO/IEC 19086-1 standard.

## Acknowledgement

This work was supported by Comunidad de Madrid (Spain) under the project S2013/ICE-3095-CM (CIBERDINE).

# Appendix A. Summary of the solutions for the security challenges in free-access public cloud storage providers

The tradeoff between security and low-cost storage solutions is a main goal of the guideline provided in this work. In specific, we have been mainly concerned with the adoption of free cloud storage services and the adequate security solutions for the sake of privacy protection and legal/normative compliance. Regarding the security solutions, it is very important which of them can be implemented by the cloud user, which of them require the use of an additional server controlled by the cloud user, and which of them cannot be deployed without the collaboration of the CP.

| Solution | Server side | Cloud User side |
|---|---|---|
| **Challenge 1: Authentication** | | |
| Password-based Authenticated Key Exchange (PAKE) | ♣ | |
| Secure Remote Password (SRP) | ♣ | |
| Two-Step Verification (2SV) | ♣ | |
| Anonymous authentication mechanisms | ♣ | |
| Cloud Access Security Broker (CASB) | ♦ | |
| Dynamic Authorisation Management (DAM) | ♦ | |
| Single Sign-On (SSO) | ♦ | |
| **Challenge 2: Information encryption** | | |
| Client-side encryption | | ✓ |
| Data fragmentation | ♦ | |
| **Challenge 3: Inappropriate modifications of information assets** | | |
| Retrievability proofs | | ✓ |
| Verification by Hash functions and digital signatures | | ✓ |
| Identity traceability mechanisms | ♣ | |
| **Challenge 4: Availability** | | |
| Multi-cloud environments | ♦ | ✓ |
| **Challenge 5: Data location** | | |
| Distance bounding protocols | | ✓ |
| **Challenge 6: Data deduplication** | | |
| Convergent Encryption (CE) + popularity of data segments | ♦ | |
| **Challenge 7: Version control of encrypted data** | | |
| Split the files uploaded to the cloud in data objects | ♣ | |
| Solutions based on server-side encryption | ♦ | |
| **Challenge 8: Assured deletion of data** | | |
| Client-side encryption | | ✓ |
| **Challenge 9: API's validation** | | |
| Secure Development Lifecycle (SDL) and automatic tools | ♣ | ✓ |
| **Challenge 10: Usable security solutions** | | |
| Security-by-design and Privacy-by-design paradigms | ♣ | ✓ |
| Usability tests | ♣ | ✓ |

Table A.1: Summary of the solutions for each security challenge in free cloud storage. In some case the deployment of solution requires a central authority, i.e., a server. This server can be the CP (♣) or other server that is not controlled by the CP that is used to store the data (♦). In some security solutions (those with a ♣/♦ in the CP side, and a ✓ in the user side) the final deployment can be only achieved through the collaboration of the CP and the cloud user.

For a summary of the different solutions and the entity implied in its implementation, see the Table A.1. In this table we can observe that there are some solutions that can be implemented either on the CP side or in the user side, while others (as the validation of APIs) demand the collaboration between the CP and the cloud user. In addition, in Table B.1 we enclose a summary of the security solutions for each challenge, along with the corresponding limitations and the main references to tackle the underline problem.

| Challenge | Solutions | Limitations | References |
|---|---|---|---|
| Authentication | PAKE; SRP; 2SV; Anonymous authentication mechanisms; CASB; DAM; SSO | Poor implementation of the cryptographic functionalities; Difficult combination of usability goals and security concerns | [20, 21, 23, 24, 25, 27, 30, 32] |
| Information encryption | Client-side encryption; Data fragmentation | Key management: MPC systems; Loss of functionalities: homomorphic encryption as solution; Sharing of encrypted files: digital envelope container as solution | [41, 56, 46] |
| Inappropriate modifications of assets | Hash functions; Retrievability proofs; Digital signatures; Identity traceability mechanisms | Performance | [45, 60, 61] |
| Availability | Multi-cloud environments | Trust in third parties in multi-cloud solutions | [69, 70] |
| Data location | Distance bounding protocols | Solutions dependent on the storage service and the Internet communication | [72, 73] |
| Data deduplication | Convergent Encryption (CE) + popularity of data segments; ClouDedup approach | CE suffers of several weaknesses: confirmation of a file attack, learn the remaining information attack | [76, 108, 109] |
| Version control of encrypted data | Split the files uploaded to the cloud in data objects; SparkleShare; Git-crypt | High management costs for the user | [83, 84, 85, 86] |
| Assured deletion of data | Client-side encryption | Key management; Data encryption does not protect against changes in the users groups | [83] |
| API's validation | Use of a Secure Development Lifecycle (SDL) and automatic tools for software validation | The human factor | [8, 93, 102, 101, 110, 111] |
| Usable security solutions | Security-by-design and Privacy-by-design paradigms; Usability tests | Security and privacy are rarely the user's main goal but these solutions must be user friendly | [88, 91, 92, 94, 93, 8] |

Table A.2: Summary of security challenges and solutions in the public cloud storage.

# Appendix B. Summary of the main characteristics of the most popular free cloud storage services

In this section we are going to compare the most relevant free cloud storage services, starting with the Table B.1, which compares the main features of them.

|  | **OneDrive** | **Dropbox** | **Google Drive** | **Box** |
|---|---|---|---|---|
| **Best for** | Devoted Windows Users | Lightweight Users | Teams and Collaboration | Enterprise Solutions |
| **Area of specialization** | Collaboration; Microsoft Office 365 Included With 1TB Purchase | Compatibility With Other Services | Collaboration | Compatibility With Other Services; Business Use |
| **File size restrictions?** | 10GB | 10GB with website, none with Dropbox apps | 5TB | 250MB for free plan, 5GB for paid personal plan |
| **Free storage?** | 5GB | 2GB | 15GB | 10GB |
| **Can I earn extra fee storage?** | No | Yes | No | No |
| **Cheapest Premium Option** | $1.99/month for 50GB | $10/month for 1TB | $2/month 100GB, $10/month for 1TB | $10/month for 100GB |
| **File versioning** | Yes | Yes | Yes | Yes |
| **OSes supported** | Windows, Mac, Android, iOS, Windows Phone | Windows, Mac, Linux, Android, iOS, Windows Phone, BlackBerry, Kindle Fire | Windows, Mac, Android, iOS | Windows, Mac, Android, iOS, Windows Phone, BlackBerry |
| **Server Location** | Worldwide | United States | Worldwide | Worldwide |
| **iOS App User Rating** | 4 | 3.5 | 4.5 | 4 |
| **Android App User Rating** | 4.4 | 4.4 | 4.3 | 4.2 |
| **Windows App User Rating** | 4.2 | 3.5 | 3.9 | 4.4 |

Table B.1: Main features of OneDrive, Dropbox, Google Drive and Box [112, 103, 113].

*Appendix B.1. OneDrive*

Previously known as SkyDrive, OneDrive is Microsoft's solution to offer simultaneously a cloud storage service and the Office suite [103], being the best option for many Windows users [104]. Its biggest strength is that it works closely with Microsoft Office apps, such as Word or PowerPoint, so when you launch one of those applications you will see a list of recent documents saved to OneDrive [114]. If you have an Office 365 subscription and you open a document saved in OneDrive, you can collaborate on it in real time with other people [112].

It is remarkable that if privacy is a major concern, regarding privacy, we should note that Microsoft reserves the right to scan users' files to discard copyright infringement or other illegal contents [104]. Besides, data are encrypted in transit using SSL, but it remains unencrypted at rest if you are using a free account. On the other hand, if you are a business OneDrive user, you can have an individual encryption of each file that you have in your account. Each file is encrypted with a unique key, so if you lose this key, only one file will be compromised. In addition, all OneDrive users can access with a two-step verification mechanism, which further protects the login via a One Time Code app or a text message [27].

Table B.2 shows the main advantages and disadvantages of using OneDrive:

| Pros | Cons |
| --- | --- |
| It is installed by default on Windows systems | You are required to sign up for a Microsoft account |
| Users get full access for free and it is easy to open and edit files from OneDrive in Microsoft's other applications, such as Word or Excel | File management is less intuitive than that of its competitors |
| Both iOS and Android interfaces are simple and clean | OneDrive's automatic file organisation does not always put files in the correct folders |
| Signing up for OneDrive gets you a Microsoft account, which gives you access to Outlook, Xbox Live and other Microsoft Services | |

Table B.2: Advantages and disadvantages of OneDrive [112, 103]

Therefore, if you have a Windows PC, tablet and phone, and you need to get to your files from any device with little effort, OneDrive is probably your best option [112].

*Appendix B.2. Dropbox*

Dropbox is the most well-known cloud storage service. Established in 2008, it currently has more than 500 million users worldwide [115]. Its main feature is simply how well established and easy to use it is, and it is reliable and simple to set up.

Dropbox offers the least amount of initial free storage at 2GB, but this can be expanded to 16GB through referrals (getting others to sign up) [116]. Then, if you only want to store some documents, this free storage will be enough for you. The problem will begin if you want to store any kind of media.

In order to store a file in Dropbox, you can do it through the website or the desktop app. The latter works by creating a local folder on your device or PC that then syncs with an online version. This means you have all your data available whether you are on or offline. This does not apply to mobile devices [104]. This desktop app lives in your file system so that you can easily move files from your computer to the cloud and vice versa by dragging and dropping them into your Dropbox folder. The service automatically and quickly syncs your files across all of your devices, so you can access everything, everywhere [112].

Folders and files can also be shared with others but you cannot set permissions on the Basic account, so files can be edited (and even deleted) by other users. However, with this Basic account you can back up any changes to files for 30 days [104].

Related to security features, Dropbox uses SSL for data in transit and AES-256 for data at rest [27]. Dropbox also includes two-step authentication [104], and its employees cannot view the content of the files you store but can access metadata if they need to in order to provide tech support. Only a small number of employees can access stored files if required to for legal reasons [27]. Furthermore, lost or stolen devices can be easily 'unlinked' from your account to further mitigate the risk of unauthorised access.

The business version, Dropbox Pro, adds an ability to enable viewer permissions for collaborative usage and set both passwords and expirations for shared links [27].

In Table B.3 there are the main advantages and disadvantages of using Dropbox:

| Pros | Cons |
|---|---|
| It performs equally fantastic on all major operating systems and devices | The freemium version is 2GB |
| The free plan includes a version history | The only paid option (Dropbox Pro) is $10/month for 1TB |
| Uploaded files have a 256-bit AES encryption and are protected by SSL/TLS in transit | |
| There is no file-size limit | |
| The service is so simple and elegantly designed, that it is easy for anyone to master | |
| Its desktop applications seamlessly blend with your computer's file system | |

Table B.3: Advantages and disadvantages of Dropbox [112, 103]

Therefore, Dropbox is the best option if you are new to cloud storage. This is due to file management is intuitive, and all the apps (including the browser client) are built around a minimalistic theme that offers the same fluid experience on all major operating systems and devices [103].

*Appendix B.3. Google Drive*

Google combines a complete set of office tools with cloud storage in Google Drive, and it is mainly designed to create and share documents, working in a collaborative way. You get a little bit of everything with this service, including a word processor, spreadsheet application, and presentation builder, plus 15GB of free storage space [112]. It is remarkable that the storage space is shared across your Gmail account, photos you upload to Google+ and any documents you create in Google Drive, so if you have large attachments on emails then they will count in these 15GB [104]. This cloud service allows you to preview attachments from Gmail in Google Drive, and save those files to your cloud [112], and to quickly attach documents from Google Drive [116].

Google Drive works in the same way as most cloud storage solutions, with a local folder on your PC linked to a duplicate cloud version. Clients are available on PC and Mac, with mobile versions for Android and iOS. Versioning is supported, as real-time collaboration on documents via the Google Docs app [104].

Data stored on Google Drive is encrypted in 128-bit AES rather than the 256-bit employed by Box, OneDrive, and Dropbox [104]. Data in transit is protected using SSL, and also implements some "internal measures" to look out for potential compromised account login activity [27].

Google claims that they do not look into the content of your Google Drive folder unless compelled by law enforcement agencies. Like other cloud storage services, you can set up two-step verification on your account to add another layer of security [104].

In Table B.4 there are the main advantages and disadvantages of using Google Drive:

| Pros | Cons |
|---|---|
| It includes a top-notch office suite | Drive requires a Google account, which means you inherit your own Gmail address and Google+ profile. If you are worried about Google's reach into your private life, this might be a deal breaker |
| It offers extensive file-sharing and synchronous-collaboration features | If you use Google Drive's tools to create documents, spreadsheets or presentations, you must export those files to edit them in another program |
| The search function utilizes Google's image-recognition technology | You have to share your storage space with Gmail |
| Google Drive requires very little setup if you already have a Google account | |
| If you use Gmail, it is easy to save attachments from your e-mail directly to Drive with just a few clicks | |
| The app can automatically back up your photos on its own, without the need for the separate Google Photos app | |

Table B.4: Advantages and disadvantages of Google Drive [112, 103]

Google Drive is the best option for Google's users, since most of them have already an account on Gmail, so they could get the most with it using the tools Google Drive offers in the cloud [112].

*Appendix B.4. Box*

This cloud storage solution is the best option for many companies, since Box let them share files with colleagues, assign tasks, leave comments on someone's work, and get notifications when a file changes [112].

Like other cloud storage services, Box has a desktop app which allows sync your files between your hard drive and the cloud. If you work with documents and spreadsheets, you will not have storage problems with Box, since the user has to keep in mind the fact that Box limits files to 250MB. The problem will come if you for example want to store some huge videos [112].

Another feature it is really useful for companies is that Box gives them a lot of control over the privacy of their files, so they can decide who in their business can view and open specific folders and files, as well as who can edit and upload documents. They can even password-protect individual files and set expiration dates for shared folders. Box protects the confidentiality and integrity of your files: data in transit is protected with TLS and data at rest with 256-bit AES. Encryption keys are securely stored in separate locations [117].

In Table B.5 there are the main advantages and disadvantages of using Box:

| Pros | Cons |
| --- | --- |
| A huge selection of apps and services offer Box integration | The free account has a 250MB file-size limit |
| It is basically a lightweight business toolkit. Box comes with tons of tools for businesses, including collaboration and file privacy control | Automatic photo uploading requires a Pro account |
| The free plan includes 10GB of space | It supports a limited variety of file type |
| | The service's endless list of sharing and privacy features can be lost on someone who is just using the service for personal storage |
| | Because of all those features, it can feel overwhelming to navigate the Box website if you are only trying to manage a few files and folders |

Table B.5: Advantages and disadvantages of Box [112, 103]

We consider that Box is the best cloud storage service for teams of employees working together on projects, and for large companies that need a place to securely share documents with everyone [112].

*Appendix B.5. Other cloud storage solutions*

Nowadays, there are some cloud solutions which are more focused on trying to solve the security problems which threaten cloud storage environments. In fact, some of them are really well-known, like Cloudfogger[19], Boxcryptor[20], Sookasa[21], EnCifra[22] and ScapePrivacy[23]. CipherCloud[24] is another platform that serves as a gateway for data encryption in real-time and before sending them to a cloud environment. They emphasise that the encryption keys are stored locally, so they are not shared with the CP. However, the main problem in all of these implementations is that they are private solutions. To begin with, the user has to sign up in a service which probably stores her authentication data. Therefore, these solutions have a server in which the user should not trust if she wants to have the control over her data. Moreover, in these types of services it is even more difficult to verify the security requirements than in Open Source projects, because in this last case the source code is available for the whole community.

---

[19] www.cloudfogger.com Accessed 2016-06-04
[20] www.boxcryptor.com Accessed 2016-06-04
[21] www.sookasa.com Accessed 2016-06-04
[22] www.encifra.net Accessed 2016-06-04

Another well-known solution is Mega, which is a relatively new cloud storage option in the industry compared to the more established solutions such as Google Drive, OneDrive and Dropbox. Mega also has its own local clients for Windows, OS X and Linux, but its main features are the 50GB of free storage space that it provides, and its high level of security, since it offers client-side encryption of data which means files are encrypted before they are uploaded to the server. In fact, Mega claims that it does not have any way of accessing user's information, as Mega allows the user to hold her master decryption key. It means that anything the user stores in her account is only accessible by her. The main problem for the user comes in the context of key management, since if the user loses her master key, there is no way to recover her account, so her files will remain encrypted and stored on the server [118]. However, as it has remarked before, the user should not trust this CP, as we can conclude from the analysis of Mega by the MEGApwn project [119]. In this study it is shown that the encryption master key in Mega was not actually encrypted and could be retrieved by Mega or anyone else with access to the user's computer.

Currently, there are two solutions which do not rely on external servers for the key generation task. The first one is OmniShare [120], which protects cloud users with client-side encryption with strong keys without significant reduction in the user experience. This solution is also available on different types of devices (mobiles, desktop). However, its key distribution protocol requires that these devices have to be close in order to make the key exchange. The other solution is Cryptomator[25], a free client-side encryption for cloud files which does not have a key distribution protocol for sharing encrypted files: the user is in charge of distributing the encryption key.

---

[23] www.scapeprivacy.com Accessed 2016-06-04
[24] www.ciphercloud.com Accessed 2016-06-04
[25] www.cryptomator.org Accessed 2016-06-04